\documentclass[journal]{IEEEtran}
\usepackage{graphicx}
\usepackage[right=0.57in, left=0.57in, top=1.1in, bottom=0.7in]{geometry}
\usepackage[center]{caption}
\usepackage{subcaption}
\usepackage{cite}
\usepackage{color}
\usepackage{amsmath,amssymb,amsthm}

\newcommand\norm[1]{\left\lVert#1\right\rVert}
\usepackage{algorithm,algorithmic,multicol}

\begin{document}
\title{Covariance-Based Hybrid Beamforming for Spectrally Efficient Joint Radar-Communications}

\author{\IEEEauthorblockN{Evangelos Vlachos$^1$ and Aryan Kaushik$^2$} 
\\
\IEEEauthorblockA{
$^1$Industrial Systems Institute, ATHENA Research and Innovation Centre, Greece.\\
$^2$School of Engineering and Informatics, University of Sussex, United Kingdom. \\
Emails: evlachos@athenarc.gr, aryan.kaushik@sussex.ac.uk}}

\maketitle

\begin{abstract}
Joint radar-communications (JRC) is considered to be a vital technology in deploying the next generation systems, since its useful in decongestion of the radio frequency (RF) spectrum and utilising the same hardware resources for dual functions. Using JRC systems for dual function generates interference between both the operations which needs to be addressed in future standardization. Furthermore, JRC systems can be advanced by deploying hybrid beamforming which implements fewer number of RF chains than the number of transmit antennas. This paper designs a robust hybrid beamformer for minimizing the interference of a JRC transmitter via RF chain selection resulting into mutual information maximization. We consider a weighted mutual information for the dual function JRC system and implement a common analog beamformer for both the operations. The mutual information maximization problem is formulated which is non-convex and difficult to solve. The problem is simplified to convex form and solved using Dinkelbach approximation abased fractional programming. The performance of the optimal RF selection based proposed approach is evaluated, compared with baselines and its effectiveness is inferred via numerical results.
\end{abstract}


%
\IEEEpeerreviewmaketitle

\section{Introduction}
It is expected that the fifth generation (5G) connections will constitute towards 1.4 billion mobile devices by 2023 \cite{cisco2020}, and there will be 4.4 billion 5G subscriptions worldwide presenting faster growth than previous generation standards \cite{ericsson2022}. The ever-increasing growth and user demand in sixth generation (6G) systems will lead to service providers finding advanced solutions to make use of available resources for maximum capacity possible. Integrated sensing and communication (ISAC) technology provides a potential solution towards efficient use of hardware and spectral resources, while decongesting the crowded sub-6 GHz spectrum used for most of the mobile communications \cite{zhangJSTSP2021, liuTCOM2020}. In ISAC systems, sensing collects and extracts information, such as target detection in the case of radar sensing, and using radio waves for tracking movement, while communication possesses transfer of information.

In the scope of ISAC, there are emerging joint radar and communications (JRC) systems which accommodate multiple functionalities including wireless communications and radar sensing, and resulting into wide applications for future wireless, defence, aerial, space, vehicular and internet-of-intelligent-things etc. \cite{zhangJSTSP2021, liuTCOM2020, cuiNet2021, aryanIET2022}. The JRC systems are classified into three categories \cite{zhangJSTSP2021}: i) \emph{communication-centric} JRC implementing radar sensing as secondary, ii) \emph{radar-centric} JRC integrating wireless communication as the secondary, and
iii) \emph{joint JRC} design that offers tunable trade-off between both the operations. For joint JRC designs, there have been recent technical advancements in terms of achieving energy and hardware efficient systems \cite{dizdarOJCOMS2022, aryanICC2021, aryanICC2022}. The JRC systems have been also investigated for optimal transmit waveform design with hardware efficient setups using low complexity analog architecture \cite{aryanJCNS2022}, and information embedding approaches for improved performance \cite{shenGCOM2022}.

Besides energy and hardware efficient designs of JRC systems, there has been recent attention on spectral efficiency maximization such as in \cite{aryanIET2022} where interference and hardware distortion parameters are taken into account. There is indeed requirement of also observing trade-off between the radar mutual information and communication rate in mutual information of JRC systems \cite{aryanICC2022}. 
Furthermore, hybrid beamforming based energy and hardware efficient designs incorporated with massive multiple-input multiple-output (MIMO) antenna setup can effectively achieve high degrees-of-freedom while achieving low-cost outcomes. 
However, the use of hyrbid beamforming has not been widely explored for JRC systems in the context of mutual information maximization. The joint design of JRC using a weighted mutual information while reduce the interference occurring from each of the operations, can provide further benefits in terms of effectiveness of such system. This urges researchers to look into more robust hybrid beamformers that take into account interference terms and achieve a spectrally efficient JRC system.

\subsubsection*{Contributions} 
In this paper, we design a covariance-based robust hybrid beamformer for dual function JRC transmission that maximizes the mutual information via RF chain selection and minimizing the interference from one operation on the other. JRC transmission employs a simplified structure with common analog beamformer for both the operations. A joint mutual information maximization problem with weighting is formulated which is non-convex and difficult to solve. The optimization problem is approximated to convex form for good performance using Dinklebach approximation based fractional programming. It is important to note that the proposed technique uses only the channel second order statistics and not instantaneous channel which makes it more robust to channel estimation errors. The numerical results are presented for the proposed method which is compared with several baselines and for different weighting factor values, to show the effectiveness of the proposed optimal RF selection based approach.

\emph{Notation:} The notation \textbf{M} represents matrix, \textbf{m} represents vector, m is scalar entity, tr(.) and $|.|$ represent trace and determinant functions, $(.)^\text{T}$ denotes transpose, $(.)^\text{H}$, $\norm{.}_F$ and $\Vert.\Vert_2$ represent complex conjugate transpose, Frobenius norm, and the Euclidean norm, respectively. The notation $[\textbf{M}]_{m,k}$ is $(m,k)$-th element in matrix $\mathbf{M}$ and $\mathbf{m}_m$ is $m$-th element of $\mathbf{m}$; $\textbf{I}_{J}$ is $J$-size identity matrix; $\mathbf{I}_{J \times K}$ denotes column concatenated matrix as $[\mathbf{I}_J \,\, \mathbf{0}_{J \times K}]$. $\mathbb{C}$ and $\mathbb{R}$ represent sets of complex and real numbers, respectively, and $\mathcal{E}$ denotes expectation operator. The notation $\mathcal{C}\mathcal{N} (m, n)$ denotes complex Gaussian vector which has mean $m$ and variance $n$.

\section{System and Channel Models}
The considered system is composed of $N$ antenna terminals at both transmitter and receiver, which assists the downlink communication between a base-station/transmitter and $K$ users (UE)/receiver, as shown in Fig. 1. The transmitter implements a dual function JRC with hybrid beamforming which employs baseband precoding and analog precoding. Note that the analog precoder $\mathbf{F}_\text{RF}$ is joint for both radar-communications operation. The transmit signal constitutes $\mathbf{s} = \left[ \begin{array}{cc} \mathbf{s}_\text{C}^T & \mathbf{s}_\text{R}^T \end{array}\right]^T$, where $\mathbf{s}_\text{C} \in \mathbb{C}^{\frac{N}{2} \times 1}$ refers to the communication vector term and $\mathbf{s}_\text{R} \in \mathbb{C}^{\frac{N}{2} \times 1}$ is for the radar operation.
The transmit signal $\mathbf{x}$ for JRC is,
\begin{equation}
\mathbf{x} = \mathbf{F}_\text{RF} \mathbf{D} \mathbf{s} = \mathbf{F}_\text{RF} (\mathbf{D}_\text{C} \mathbf{s}_\text{C} + \mathbf{D}_\text{R} \mathbf{s}_\text{R}) 
\end{equation}
where $\mathbf{D}^{N \times N}$ is a diagonal matrix with binary entries, representing the state of the switches; $\mathbf{F}_\text{RF} \in \mathcal{F}^{N \times N}$ consists of a network of phase shifters, where the elements of analog precoder with constant-modulus entries.

The received time-domain baseband signal after the analog combiner is represented as
\begin{align}
    \mathbf{y} &= \mathbf{F}_\text{RF}^\text{H} \mathbf{x} = \mathbf{F}_\text{RF}^\text{H} \mathbf{H}\mathbf{F}_\text{RF} \mathbf{D} \mathbf{s} \\
    & =  \mathbf{F}_\text{RF}^\text{H} \mathbf{H}\mathbf{F}_\text{RF} (\mathbf{D}_\text{C} \mathbf{s}_\text{C} + \mathbf{D}_\text{R} \mathbf{s}_\text{R}) \\
    &= \underbrace{\mathbf{F}_\text{RF}^\text{H} \mathbf{H}\mathbf{F}_\text{RF} \mathbf{D}_\text{C} \mathbf{s}_\text{C}}_{\triangleq \mathbf{y}_\text{C}} + \underbrace{\mathbf{F}_\text{RF}^\text{H} \mathbf{H}\mathbf{F}_\text{RF} 
 \mathbf{D}_\text{R} \mathbf{s}_\text{R}}_{\triangleq \mathbf{y}_\text{R}},
\end{align}
where $\mathbf{H} \in \mathbb{C}^{N \times N}$ represents the channel term.
\begin{figure}
 \begin{center}
 	\includegraphics[width=0.5\textwidth, trim=50 100 340 140,clip]{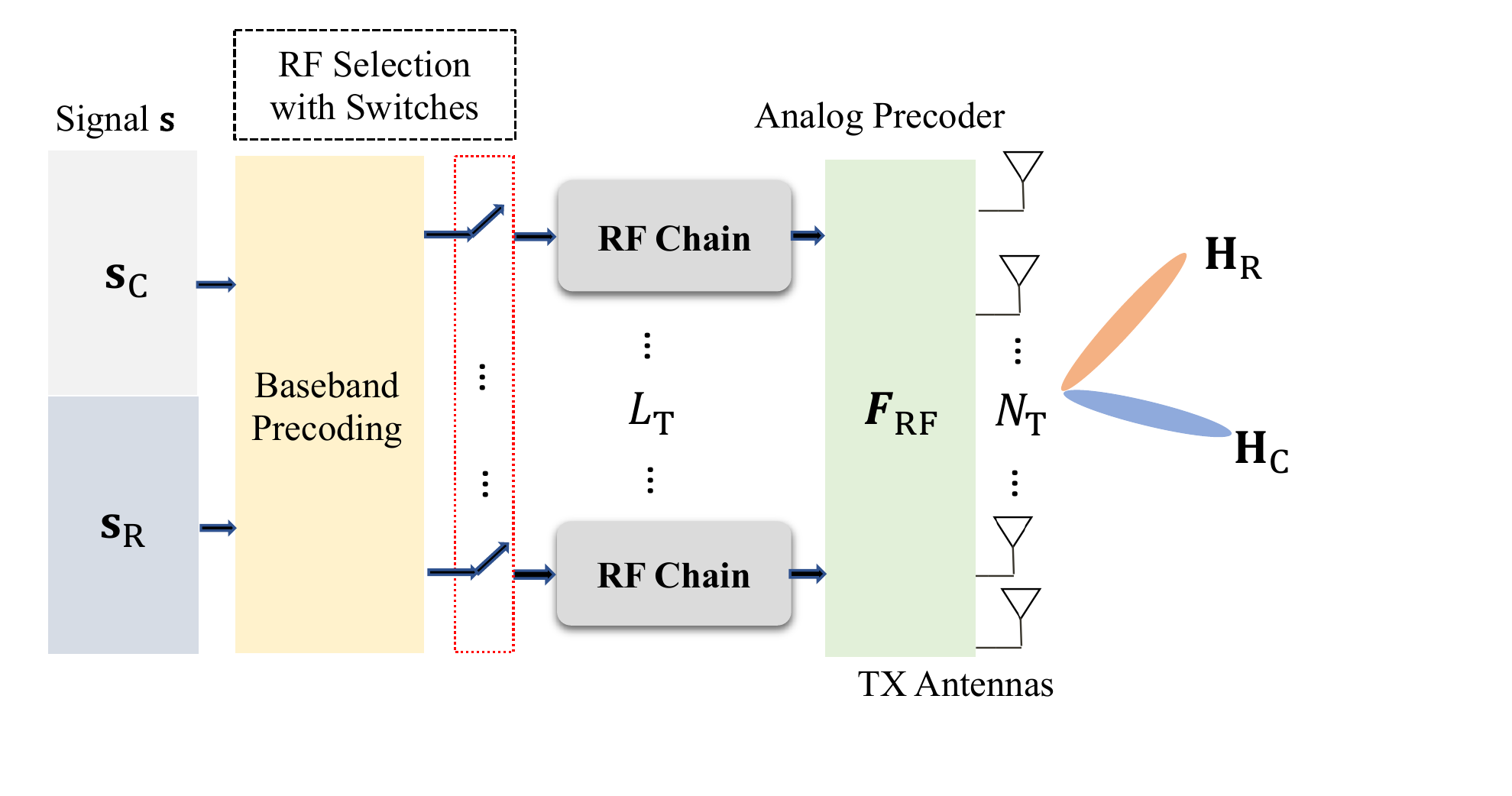}
 \end{center}
 \vspace{-7mm}
 	\caption{Robust hybrid beamforming for JRC transmission.}
 \vspace{-5mm}
\end{figure}

Following that, we describe the general channel model that can be used for both communication and radar operations. For $N_c$ multipaths, the baseband channel matrix is represented as
\begin{equation}\label{eq:channel_model}
\mathbf{H} = \frac{N}{\sqrt{N_\textrm{c}}} \sum_{l=1}^{N_{\textrm{c}}}  \alpha_{l} \mathbf{a}_{\textrm{R}}(\phi_{l}^{r}) \mathbf{a}_{\textrm{T}}^H(\phi_{l}^{t}),
\end{equation}
with $\alpha_{l}$ being the channel gain of $l$-th path, and $N_\textrm{c}$ being the number of clustered multipaths. The term $\textbf{a}_{\textrm{T}}(\phi_{l}^{t}) = \frac{1}{\sqrt{N}}{[1, e^{j \frac{2 \pi}{\lambda}d\sin(\phi_{l}^{t})}, ..., e^{j (N_{T}-1)\frac{2 \pi}{\lambda}d\sin(\phi_{l}^{t})}]}^{T}$ represents the steering vector for transmission. The departure angle is represented as $\phi_{l}^{t}$ with $d$ being spacing between antennas and $\lambda$ being wavelength. Similarly, the term $\textbf{a}_{\textrm{R}}(\phi_{l}^{r})$ represents array response vector for receiver. The angle of arrival is represented as $\phi_{l}^{r}$. It is assumed that the channel state information (CSI) of communication channel is known at transmitter and UE sides. Furthermore, the antenna setup in \eqref{eq:channel_model} follows an uniform linear array (ULA) configuration while the proposed method is independent of antenna configuration, for instance, it can also be applied to circular array and rectangular array configurations.

\vspace{-3mm}
\section{JRC Mutual Information Maximization}
\subsection{Mutual Information}

In this section we present mutual information expressions for the dual function JRC system model where we consider two of the possible cases: (i) with no impact of interference of one operation over the other operation which presents an idealized case, and (ii) while taking into account the impact of interference of one operation over the other operation. The communication spectral efficiency is well known in literature. 

In terms of the literature surrounding radar mutual information, reference \cite{35} describes waveform design approach while taking into account radar mutual information between target reflections and target responses. Furthermore, the main purpose of the radar mutual information is to evaluate the radar performance, such as in \cite{36}, \cite{37} mutual information is described between target impulse response and echo/reflected signal, provided there is prior knowledge of the dual function JRC transmit signal or radar-only transmit signal. JRC systems such as in \cite{38}, express the radar mutual information between the signal and target impulse response of radar and communication signals, while assuming that both radar and communication signals partly have common information on the target. Next, we consider joint channel capacity terms for JRC with no interference (of one operation over the other operation) and with interference scenarios.

\subsubsection{Separate subsystems} Let us first consider the idealized case where the two systems have dedicated separate hardware that do not interfere with each other. The mutual information (MUI) can be expressed as \cite{tse2005fundamentals}:
\begin{equation}\label{eq:MUI_optimal}
     I^\text{opt} = \log_2 \vert \mathbf{I} + \frac{1}{\sigma_n^2} \mathbf{H}_\text{C} \mathbf{\Sigma}_\text{C} \mathbf{H}_\text{C}^\text{H} \vert +  \log_2 \vert \mathbf{I} + \frac{1}{\sigma_n^2} \mathbf{H}_\text{R} \mathbf{\Sigma}_\text{R} \mathbf{H}_\text{R}^\text{H} \vert,
\end{equation}
where $\mathbf{\Sigma}_\text{C}$, $\mathbf{\Sigma}_\text{R}$ are the covariance matrices of the precoder outputs with $\Vert \mathbf{\Sigma}_\text{C} \Vert^2 = \Vert \mathbf{\Sigma}_\text{R} \Vert^2 = P_\text{max}/2$.  The term $\sigma_n^2$ accounts for the noise variance.

\subsubsection{Joint system for both operations} The MUI for this case also includes the interference terms between the subsystems:
\begin{align}
I^\text{jnt} = &\log_2 \vert \mathbf{I} + \frac{1}{\sigma_n^2 + \sigma_\text{R}^2}  \mathbf{H}_\text{C}^\text{H} \mathbf{\Sigma}_\text{C} \mathbf{H}_\text{C} \vert \nonumber \\ &+  \log_2 \vert \mathbf{I} + \frac{1}{\sigma_n^2 + \sigma_\text{C}^2} \mathbf{H}_\text{R}^\text{H} \mathbf{\Sigma}_\text{R} \mathbf{H}_\text{R} \vert \le I^\text{opt}, \label{eq:I_jnt}
\end{align}
where $\sigma_\text{R}^2$ corresponds to the radar interference while communication operation is taking place in the dual function JRC system, and similarly $\sigma_\text{C}^2$ corresponds to the communication interference to the radar operation, with $\sigma_\text{R}^2 = \Vert \mathbf{H}_\text{R} \mathbf{\Sigma}_\text{C}^{1/2}\Vert^2$ and $\sigma_\text{C}^2 = \Vert \mathbf{H}_\text{C} \mathbf{\Sigma}_\text{R}^{1/2} \Vert^2$, respectively.

\vspace{-3mm}
\subsection{Problem Formulation}
Using the \eqref{eq:I_jnt}, the problem of precoder design for maximizing the mutual information is expressed as:
\begin{align}
    \max_{\mathbf{\Sigma}_\text{C}, \mathbf{\Sigma}_\text{R}} & \log_2 \vert \mathbf{I} + \frac{1}{\sigma_n^2 + \sigma_\text{R}^2}  \mathbf{H}_\text{C} \mathbf{\Sigma}_\text{C} \mathbf{H}_\text{C}^\text{H} \vert \nonumber \\ & + \log_2 \vert \mathbf{I} + \frac{1}{\sigma_n^2 + \sigma_\text{C}^2} \mathbf{H}_\text{R} \mathbf{\Sigma}_\text{R} \mathbf{H}_\text{R}^\text{H} \vert \nonumber \\
    \text{subject to } & \Vert \mathbf{\Sigma}_\text{C} \Vert^2 + \Vert \mathbf{\Sigma}_\text{R} \Vert^2 = P_\text{max}.
\end{align}
In this idealized case, the precoders can be designed so as to nullify the interference of each other. Specifically, by setting 
\begin{align}
\mathbf{\Sigma}_\text{C}^{1/2} &= \text{nullspace}(\mathbf{H}_\text{R})\textrm{, and} \label{eq:precoder_9}\\ 
\mathbf{\Sigma}_\text{R}^{1/2} &= \text{nullspace}(\mathbf{H}_\text{C}),\label{eq:precoder_10}
\end{align}
then $
\mathbf{H}_\text{R}\mathbf{\Sigma}_\text{C}^{1/2} = \mathbf{H}_\text{C}\mathbf{\Sigma}_\text{R}^{1/2} = \mathbf{0}$, while 
$$
\mathbf{\Sigma}_\text{C}^{1/2} (\mathbf{\Sigma}_\text{C}^{1/2})^\text{H} =\mathbf{\Sigma}_\text{R}^{1/2} (\mathbf{\Sigma}_\text{R}^{1/2})^\text{H} = \mathbf{I}.
$$ 
Therefore, with perfect CSI knowledge for both communications and radar we have: $I^\text{jnt} = I^\text{opt}$. This case requires digital beamforming, so as to implement the precoders in \eqref{eq:precoder_9} and \eqref{eq:precoder_10}. 

Note that, perfect CSI via channel matrix recovery is difficult to be acquired, especially in the case of radars. However, a rough knowledge of the targets positions could be utilized to construct an estimation of the channel matrix for the radar case. Specifically, exploiting the sparsity property of the channel matrix in the beamspace domain, the positions of the non-zero entries are related with the AoAs and therefore the positions of the targets and the users. Given that the dimentions of the MIMO array permits the distinction between the positions, then, we could nullify each other in the beamspace domain. 

On this premise, let $\mathbf{\Omega}_\text{C}, \mathbf{\Omega}_\text{R} \in \{0,1\}^{N \times N}$ denote the binary matrices with ones at the indices that correspond to the non-zero values of the communications and radar channel beamspaces respecitvely, i.e., $\mathbf{F}^\text{H} \mathbf{H}_\text{C} \mathbf{F}$ and $\mathbf{F}^\text{H} \mathbf{H}_\text{R} \mathbf{F}$. In principle, this is related with the geometric properties of the mmWave channels, where the propagation paths correspond to groups of non-zero entries for the beamspace channel. Thus,
\begin{equation}
    \mathbf{\Omega}_\text{C} \circ (\mathbf{F}^\text{H} \mathbf{H}_\text{R} \mathbf{F}) = \mathbf{\Omega}_\text{R} \circ (\mathbf{F}^\text{H}\mathbf{H}_\text{C} \mathbf{F}) = \mathbf{0},
\end{equation}
while $\mathbf{\Omega}_\text{C} \circ (\mathbf{F}^\text{H}\mathbf{H} \mathbf{F}) = (\mathbf{F}^\text{H} \mathbf{H}_\text{C} \mathbf{F})$ and $ \mathbf{\Omega}_\text{R} \circ (\mathbf{F}^\text{H} \mathbf{H} \mathbf{F}) = (\mathbf{F}^\text{H} \mathbf{H}_\text{R} \mathbf{F})$. Therefore, the mutual information is expressed as follows:
\begin{align}
    I^{(bs)} = & \log_2 ( 1 + \frac{1}{\sigma_n^2 + \Vert \mathbf{\Omega}_\text{C} \circ (\mathbf{F}^\text{H} \mathbf{H}_\text{R} \mathbf{F}) \Vert^2 } \Vert \mathbf{\Omega}_\text{C} \circ (\mathbf{F}^\text{H}\mathbf{H}\mathbf{F}) \Vert^2\nonumber \\& 
    + \frac{1}{\sigma_n^2 + \Vert \mathbf{\Omega}_\text{R} \circ (\mathbf{F}^\text{H} \mathbf{H}_\text{C} \mathbf{F}) \Vert^2 } \Vert \mathbf{\Omega}_\text{R} \circ (\mathbf{F}^\text{H} \mathbf{H} \mathbf{F})\Vert^2).
\end{align}

\vspace{-3mm}
\subsection{Weighted MUI}
Previously, it has been proposed a weighted MUI metric that utilizes a weighting factor $\rho \in [0,1]$ for the MUI of each subsystem, i.e.,
\begin{align}\label{eq:MUI_weighting}
    I^\text{opt,w} &= 2\rho \log_2 \vert \mathbf{I} + \frac{1}{\sigma_n^2} \mathbf{H}_\text{C} \mathbf{\Sigma}_\text{C} \mathbf{H}_\text{C}^\text{H} \vert \nonumber \\ & +  2(1-\rho) \log_2 \vert \mathbf{I} + \frac{1}{\sigma_n^2} \mathbf{H}_\text{R} \mathbf{\Sigma}_\text{R} \mathbf{H}_\text{R}^\text{H} \vert,
\end{align}
and $I^\text{opt,w} = I^\text{opt}$ for $\rho = \frac{1}{2}$. Note that the JRC weighting term $\rho$ describes the dominance of one operation over the other, depending on its value between $0$ and $1$,  i.e., $\rho \in [0,1]$. For instance, when $\rho$ value is high, JRC system prioritizes communication operation and when $\rho$ value is low, JRC system prioritizes radar sensing operation. For $\rho = \frac{1}{2}$, it represents that both the operations have same weighting in dual function JRC system. Then, the problem of precoder design for maximizing the weighted mutual information is expressed as follows:
\begin{align}\label{eq:15}
    \max_{\mathbf{\Sigma}_\text{C}, \mathbf{\Sigma}_\text{R}} & 2 \rho \log_2 \vert \mathbf{I} + \frac{1}{\sigma_n^2 + \sigma_\text{R}^2}  \mathbf{H}_\text{C} \mathbf{\Sigma}_\text{C} \mathbf{H}_\text{C}^\text{H} \vert \nonumber \\ & + 2 (1-\rho) \log_2 \vert \mathbf{I} + \frac{1}{\sigma_n^2 + \sigma_\text{C}^2} \mathbf{H}_\text{R} \mathbf{\Sigma}_\text{R} \mathbf{H}_\text{R}^\text{H} \vert \nonumber \\
    \text{subject to } & 2\rho \Vert \mathbf{\Sigma}_\text{C} \Vert^2 + 2(1-\rho) \Vert \mathbf{\Sigma}_\text{R} \Vert^2 = P_\text{max}.
\end{align}
Next, we present the robust hybrid beamformer design which selects optimal RF chains for the weighted JRC system while taking into account the impact of interference. 
\vspace{-3mm}
\section{Robust Hybrid Beamformer}
Let us introduce a \textit{robust} hybrid beamforming architecture for a dual function transceiver, where the transceiver resources that correspond to the analog beamformer can be dynamically allocated for the communications and radar operations based on the weighting factor $\rho$. Specifically, given that the analog beamformer is represented by the matrix $\mathbf{F}_\text{RF} \in \mathcal{F}^{N \times N}$, $\rho$\% of its columns can be used for communications and the rest $(1-\rho)\%$ for radar. The baseband part of the hybrid beamformer is represented by $\mathbf{F}_\text{BB,} \in \mathbb{C}^{N \times N}$. Considering consecutive allocation, $[\mathbf{F}_\text{RF}]_{1:\rho N}$ for beamforming of communications, and  $[\mathbf{F}_\text{RF}]_{\rho N+1:N}$ for beamforming of radar operation, assuming that $\rho N \in \mathbb{I}$ is an integer.

Another way to represent the proposed resource allocation is by using two diagonal and binary matrices, $\mathbf{D}_\text{C}$ and $\mathbf{D}_\text{R}$ with $\text{tr}(\mathbf{D}_\text{C}) = \rho N$, and $\text{tr}(\mathbf{D}_\text{R}) = (1-\rho) N$. Thus, the hybrid beamformer for each operation can be expressed as $\mathbf{F}_\text{C} = \mathbf{F}_\text{RF} \mathbf{D}_\text{C} \mathbf{F}_\text{BB,C}$, and $\mathbf{F}_\text{R} = \mathbf{F}_\text{RF} \mathbf{D}_\text{R} \mathbf{F}_\text{BB,R}$, with $\Vert \mathbf{F}_\text{C} \Vert_F^2 = \sum_{i \in \Omega_\text{C}} [\mathbf{D}_\text{C}]_{i,i} = 2 \rho$ and $\Vert \mathbf{F}_\text{R} \Vert_F^2 =  \sum_{i \in \Omega_\text{R}} [\mathbf{D}_\text{R}]_{i,i} = 2(1-\rho)$. Note that, $\Omega_C$ and $\Omega_R$ represent the sets with indices for the non-zero entries of the beamspace channels, respectively for communications and radar.

Thus, the MUI of \eqref{eq:15} can be written as:
\begin{align}
    &\log_2 \vert \mathbf{I} + \frac{1}{\sigma_n^2} \mathbf{H}_\text{C} \mathbf{F}_\text{C} \mathbf{F}_\text{C}^\text{H} \mathbf{H}_\text{C}^\text{H} \vert +  \log_2 \vert \mathbf{I} + \frac{1}{\sigma_n^2} \mathbf{H}_\text{R} \mathbf{F}_\text{R} \mathbf{F}_\text{C}^\text{H} \mathbf{H}_\text{R}^\text{H} \vert \\
    &= \vert \mathbf{I} + 2 \rho \frac{1}{\sigma_n^2} \mathbf{H}_\text{C} \mathbf{H}_\text{C}^\text{H} \vert +  \log_2 \vert \mathbf{I} + 2 (1-\rho) \frac{1}{\sigma_n^2} \mathbf{H}_\text{R} \mathbf{H}_\text{R}^\text{H} \vert.
\end{align} 
If we consider the approximation $\log (1+x)^a \approx \log(1+a x)$, then it can be shown that:
\begin{align}
     &\vert \mathbf{I} + 2 \rho \frac{1}{\sigma_n^2} \mathbf{H}_\text{C} \mathbf{H}_\text{C}^\text{H} \vert +  \log_2 \vert \mathbf{I} + 2 (1-\rho) \frac{1}{\sigma_n^2} \mathbf{H}_\text{R}^\text{H} \mathbf{H}_\text{R} \vert \\
     &\approx 2 \rho \log_2\vert \mathbf{I} + \frac{1}{\sigma_n^2} \mathbf{H}_\text{C}^\text{H} \mathbf{H}_\text{C} \vert +  2 (1-\rho)\log_2 \vert \mathbf{I} + \frac{1}{\sigma_n^2} \mathbf{H}_\text{R}^\text{H} \mathbf{H}_\text{R} \vert \nonumber \\ 
    &= I^\text{opt,w},
\end{align}
where this approximation essentially expressed that, the weighting of the analog beamformer can be viewed as weighting of the MUI. In this manner, it is possible to define the interference between the operations as:
\begin{equation}
    \sigma_\text{R}^2 = \mathcal{E} \{ \Vert \mathbf{H}_\text{C} \mathbf{F}_\text{RF} \mathbf{D}_\text{R} \mathbf{F}_\text{BB,R}\Vert_F^2 \} = \text{tr}( \mathbf{R}_\text{C} \mathbf{F}_\text{RF} \mathbf{D}_\text{R}\mathbf{F}_\text{RF}^\text{H}),
\end{equation}
which represents the interference of the radar into the communication operation with $\mathbf{R}_\text{C} = \mathcal{E} \{ \mathbf{H}_\text{C}^\text{H} \mathbf{H}_\text{C} \}$ is the covariance matrix of the communications channel, with $\mathbf{F}_\text{BB,R}$ be an orthonormal matrix. 

Respectively,
\begin{equation}
    \sigma_\text{C}^2 = \mathcal{E} \{ \Vert \mathbf{H}_\text{R} \mathbf{F}_\text{RF} \mathbf{D}_\text{C} \mathbf{F}_\text{BB,C}\Vert_F^2 \} = \text{tr}( \mathbf{R}_\text{R} \mathbf{F}_\text{RF} \mathbf{D}_\text{C}\mathbf{F}_\text{RF}^\text{H}),
\end{equation}
which respectively represents the interference of the communications into the radar operation, with $\mathbf{R}_\text{R}$ the covariance matrix of the radar channel.

Therefore, the problem of precoder design for maximizing the mutual information becomes:
\begin{align}
    \max_{\mathbf{F}_\text{C}, \mathbf{F}_\text{R}} & 2 \rho \log_2 \vert \mathbf{I} + \frac{1}{\sigma_n^2 + \sigma_\text{R}^2}  \mathbf{H}_\text{C} \mathbf{F}_\text{C} \mathbf{F}_\text{C}^\text{H} \mathbf{H}_\text{C}^\text{H} \vert \nonumber \\ & + 2 (1-\rho) \log_2 \vert \mathbf{I} + \frac{1}{\sigma_n^2 + \sigma_\text{C}^2} \mathbf{H}_\text{R} \mathbf{F}_\text{R} \mathbf{F}_\text{R}^\text{H}\mathbf{H}_\text{R}^\text{H} \vert \nonumber \\
    \text{subject to } & 2\rho \Vert \mathbf{F}_\text{C} \Vert^2 + 2(1-\rho) \Vert \mathbf{F}_\text{R} \Vert^2 = P_\text{max}. \label{eq:prop_problem_1}
\end{align}

Considering that the RF front-end is composed by  columns of the DFT matrix, then it can be shown that there is an optimal combination of columns that belongs to the nullspace of the desired matrix. Specifically, let $\mathbf{D}_\text{C}, \mathbf{D}_\text{R} \in \mathcal{D}^{N \times N}$ denote the diagonal binary matrices, so as:
\begin{align}
    \Vert (\mathbf{F}\mathbf{D}_\text{C})^\text{H} \mathbf{H}_\text{R} \Vert_F^2 &\approx 0 \label{eq:prop_null_rad}\\
    \Vert (\mathbf{F}\mathbf{D}_\text{R})^\text{H} \mathbf{H}_\text{C} \Vert_F^2 &\approx 0 \label{eq:prop_null_com}
\end{align}
where $\mathbf{F} = e^{-j 2 \pi \mathbf{E}}$ with $\mathbf{E}=[0,1,\ldots, N-1]^\text{T} [0,1,\ldots, N-1]/N$. Furthermore, by applying the hybrid precoder to the combined channel, it nullifies the radar component. Indeed, by using the properties in \eqref{eq:prop_null_rad} and \eqref{eq:prop_null_com} we have:
\begin{align}
    (\mathbf{F} \mathbf{D}_\text{C} )^\text{H} \mathbf{H} \mathbf{H}^\text{H} \mathbf{F}\mathbf{D}_\text{C}&\approx \mathbf{F}^\text{H} \mathbf{H}_\text{C}  \mathbf{H}_\text{C}^\text{H} \mathbf{F} \\
    (\mathbf{F}\mathbf{D}_\text{R})^\text{H} \mathbf{H} \mathbf{H}^\text{H} \mathbf{F}\mathbf{D}_\text{R}&\approx \mathbf{F}^\text{H} \mathbf{H}_\text{R}  \mathbf{H}_\text{R}^\text{H} \mathbf{F}
\end{align}

\begin{figure}
    \centering
    \includegraphics[scale=0.6]{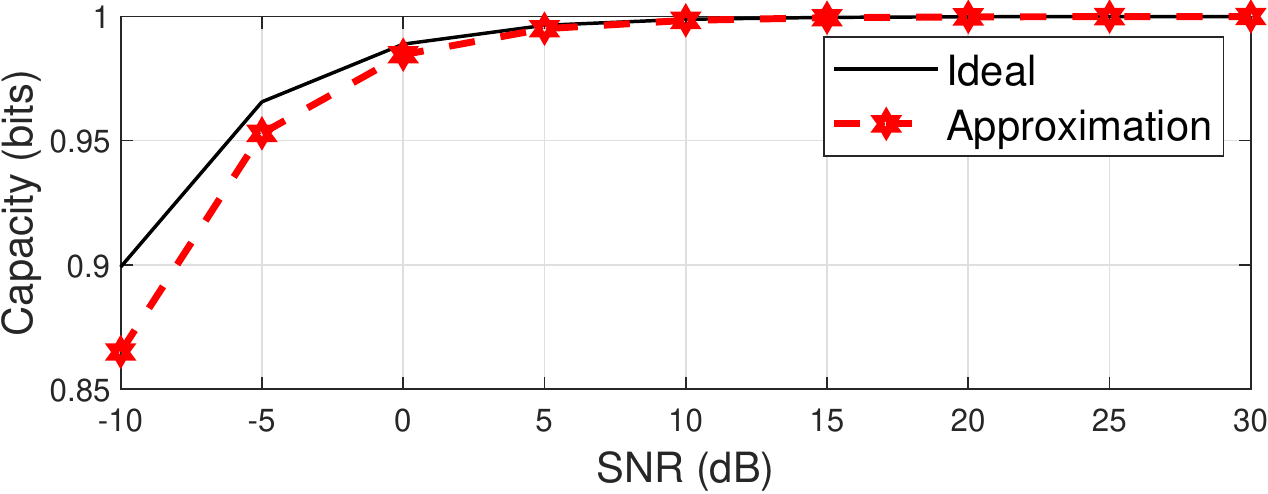}
    \vspace{-1mm}
    \caption{Ideal capacity and its approximation.}
    \label{fig:capacity_approx}
    \vspace{-3mm}
\end{figure}

The mutual information maximization problem \eqref{eq:prop_problem_1} becomes,
\begin{align} \label{eq:prop_problem_2}
    \max_{\mathbf{D}_\text{C}, \mathbf{D}_\text{R}} &\log_2 ( 1 + \frac{1}{\sigma_n^2 + \Vert (\mathbf{F} \mathbf{D}_\text{C})^\text{H} \mathbf{H}_\text{R} \Vert^2} \Vert  (\mathbf{F} \mathbf{D}_\text{C})^\text{H}\mathbf{H} \Vert^2 \nonumber \\& \hspace{3em} + \frac{1}{\sigma_n^2 + \Vert  (\mathbf{F} \mathbf{D}_\text{R})^\text{H} \mathbf{H}_\text{C} \Vert^2} \Vert  (\mathbf{F} \mathbf{D}_\text{R})^\text{H}  \mathbf{H} \Vert^2) \nonumber \\
    & \text{ subject to } \mathbf{D}_\text{C}, \mathbf{D}_\text{R} \in \{0,1\}.
\end{align}
To solve \eqref{eq:prop_problem_2} we proceed with a series of approximations which will render the problem to a convex one.
Using the first order Taylor expansion for the logarithm, $\log(1+x) \approx x$ for $x \rightarrow 0$, we have that:
\begin{align}
    &\log_2 ( 1 + \frac{1}{\sigma_n^2 + \Vert (\mathbf{F} \mathbf{D}_\text{C})^\text{H} \mathbf{H}_\text{R} \Vert^2} \Vert  (\mathbf{F} \mathbf{D}_\text{C})^\text{H}\mathbf{H} \Vert^2 \nonumber \\ & \hspace{3em} + \frac{1}{\sigma_n^2 + \Vert  (\mathbf{F} \mathbf{D}_\text{R})^\text{H} \mathbf{H}_\text{C} \Vert^2} \Vert  (\mathbf{F} \mathbf{D}_\text{R})^\text{H}  \mathbf{H} \Vert^2) \nonumber \\
    & \approx \frac{1}{\sigma_n^2 + \Vert (\mathbf{F} \mathbf{D}_\text{C})^\text{H} \mathbf{H}_\text{R} \Vert^2} \Vert  (\mathbf{F} \mathbf{D}_\text{C})^\text{H}\mathbf{H} \Vert^2 \nonumber \\ & \hspace{3em} + \frac{1}{\sigma_n^2 + \Vert  (\mathbf{F} \mathbf{D}_\text{R})^\text{H} \mathbf{H}_\text{C} \Vert^2} \Vert  (\mathbf{F} \mathbf{D}_\text{R})^\text{H}  \mathbf{H} \Vert^2 \label{eq:log_approx}
\end{align}
Indeed, in Fig. \ref{fig:capacity_approx} we depict the mutual information (capacity) between the initial (ideal) expression for the mutual information and its approximation in \eqref{eq:log_approx}, for $N=64$ and $\mathbf{D}_\text{C}=\mathbf{D}_\text{R}=\mathbf{I}_N$.

Next, the requirement for binary entries for the selection matrices is relaxed to box constraint, i.e., $[\mathbf{D}_\text{C}]_{n,n}, [\mathbf{D}_\text{R}]_{n,n} \in (0,1)$. Then, the problem \eqref{eq:prop_problem_1} can be written as:
\begin{align} \label{eq:opt_problem_prop2}
    \max_{\mathbf{D}_\text{C}, \mathbf{D}_\text{R}} & \frac{1}{\sigma_n^2 + \Vert (\mathbf{F} \mathbf{D}_\text{C})^\text{H} \mathbf{H}_\text{R} \Vert^2} \Vert  (\mathbf{F} \mathbf{D}_\text{C})^\text{H}\mathbf{H} \Vert^2 \nonumber \\& \hspace{3em} + \frac{1}{\sigma_n^2 + \Vert  (\mathbf{F} \mathbf{D}_\text{R})^\text{H} \mathbf{H}_\text{C} \Vert^2} \Vert  (\mathbf{F} \mathbf{D}_\text{R})^\text{H}  \mathbf{H} \Vert^2 \nonumber \\
    & \text{ subject to } [\mathbf{D}_\text{C}]_{n,n}, [\mathbf{D}_\text{R}]_{n,n} \in (0,1).
\end{align}

We apply the Dinkelbach approximation to the fractional problem in \eqref{eq:opt_problem_prop2}, deriving the following linear approximation:
\begin{align} \label{eq:opt_problem_prop3}
    & \max_{\mathbf{D}_\text{C}, \mathbf{D}_\text{R}} \Vert  (\mathbf{F} \mathbf{D}_\text{C})^\text{H}\mathbf{H} \Vert^2 +  \Vert  (\mathbf{F} \mathbf{D}_\text{R})^\text{H}  \mathbf{H} \Vert^2 \nonumber \\&  - 
    \kappa_\text{C}( \sigma_n^2 + \Vert (\mathbf{F} \mathbf{D}_\text{C})^\text{H} \mathbf{H}_\text{R} \Vert^2) - \kappa_\text{R}(\sigma_n^2 + \Vert  (\mathbf{F} \mathbf{D}_\text{R})^\text{H} \mathbf{H}_\text{C} \Vert^2) \nonumber \\
    & \text{ subject to } [\mathbf{D}_\text{C}]_{n,n}, [\mathbf{D}_\text{R}]_{n,n} \in (0,1).
\end{align}
Thus, we end up with the convex problem in \eqref{eq:opt_problem_prop3}, which is an approximation for \eqref{eq:prop_problem_1}, and can be solved using any available interior-point method (e.g., CVX \cite{cvx}).

The overall algorithm is summarized in \ref{algorithm:proposed}, which requires the covariance matrices for the communications and radar channels. The output are the diagonal matrices that maximize the mutual information in \eqref{eq:prop_problem_1}. Note that, we use the function $\text{thres}(\cdot)_{\rho}$, which performs thresholding on the values of its input matrix. The parameter $\rho$ determines the thresholding lower bound for the matrices $\mathbf{\tilde{D}}_\text{C}$ and $\mathbf{\tilde{D}}_\text{R}$. Next, we present the numerical results to show the significance of our proposed method.

\begin{algorithm}[t]
	\caption{Proposed Method}
	\label{algorithm:proposed}
	\begin{algorithmic}[1]
	    \REQUIRE $\mathbf{R}$, $\mathbf{R}_\text{C}$, $\mathbf{R}_\text{R}$, $\rho$
	    \ENSURE Diagonal binary matrices $\mathbf{D}_\text{C}$ and $\mathbf{D}_\text{R}$
	    \STATE Initialize $\kappa_\text{C}^{(0)} = \kappa_\text{R}^{(0)} = 1$, $\mathbf{F} = e^{-j 2 \pi \mathbf{E}}$ with $\mathbf{E}=[0,1,\ldots, N-1]^\text{T} [0,1,\ldots, N-1]/N$.
		\FOR {$i=1,2,\ldots, I_{\rm max}$}
		\STATE $c_\text{C}^{(i)} = \Vert  (\mathbf{F} \mathbf{D}_\text{C}^{(i)})^\text{H}\mathbf{H} \Vert^2$
		\STATE $c_\text{R}^{(i)} = \Vert  (\mathbf{F} \mathbf{D}_\text{R}^{(i)})^\text{H}  \mathbf{H} \Vert^2$
        \STATE $\eta_\text{C}^{(i)} = \sigma_n^2 + \Vert (\mathbf{F} \mathbf{D}_\text{C})^\text{H} \mathbf{H}_\text{R} \Vert^2$
        \STATE $\eta_\text{R}^{(i)} = \sigma_n^2 + \Vert  (\mathbf{F} \mathbf{D}_\text{R})^\text{H} \mathbf{H}_\text{C} \Vert^2$
        \STATE $\max_{\mathbf{\tilde{D}}_\text{C}} c_\text{C}^{(i)} - \kappa_\text{C}^{(i-1)} \eta_\text{C}^{(i)} \text{ s.t. } [\mathbf{D}_\text{C}]_{n,n}]_{n,n} \in (0,1)$
        \STATE $\mathbf{D}_\text{C} = \text{thres}_\rho(\mathbf{\tilde{D}}_\text{C})$
        \STATE $\max_{\mathbf{\tilde{D}}_\text{R}} c_\text{R}^{(i)} - \kappa_\text{R}^{(i-1)} \eta_\text{R}^{(i)} \text{ s.t. } [\mathbf{D}_\text{R}]_{n,n}]_{n,n} \in (0,1)$
        \STATE $\mathbf{D}_\text{R} = \text{thres}_\rho(\mathbf{\tilde{D}}_\text{R})$
        \STATE $\kappa_\text{C}^{(i)} = \frac{c_\text{C}^{(i)}}{\eta_\text{C}^{(i)}}$
        and $\kappa_\text{R}^{(i)} = \frac{c_\text{R}^{(i)}}{\eta_\text{R}^{(i)}}$
        \ENDFOR
	\end{algorithmic}
\end{algorithm}

\section{Simulation Results}
We evaluate the performance of the proposed approach to ensure the effectiveness of optimal RF selection method in a JRC system with hybrid beamforming. For evaluation of the proposed technique, we compare with the following baselines:
\begin{itemize}
    \item \textit{No interference} case, which represents the optimal upper bound for the weighted MUI, which is expressed by \eqref{eq:MUI_weighting}.
    \item \textit{With interference} case, which represents the lower bound for the weighted MUI, assuming that no measure is taken for interference mitigation.
    \item \textit{SVD nulling} is also an idealized case where digital beamforming is performed, with perfect channel knowledge for both communications and radar.
    \item \textit{Beamspace nulling} represents a hybrid beamforming technique, with perfect channel knowledge.
\end{itemize}
In the following results, for simplicity, we assume that we have the same number of users and targets, and their placement is equispaced and alternatively.

\begin{figure}[t]
    \centering
    \includegraphics[scale=0.6]{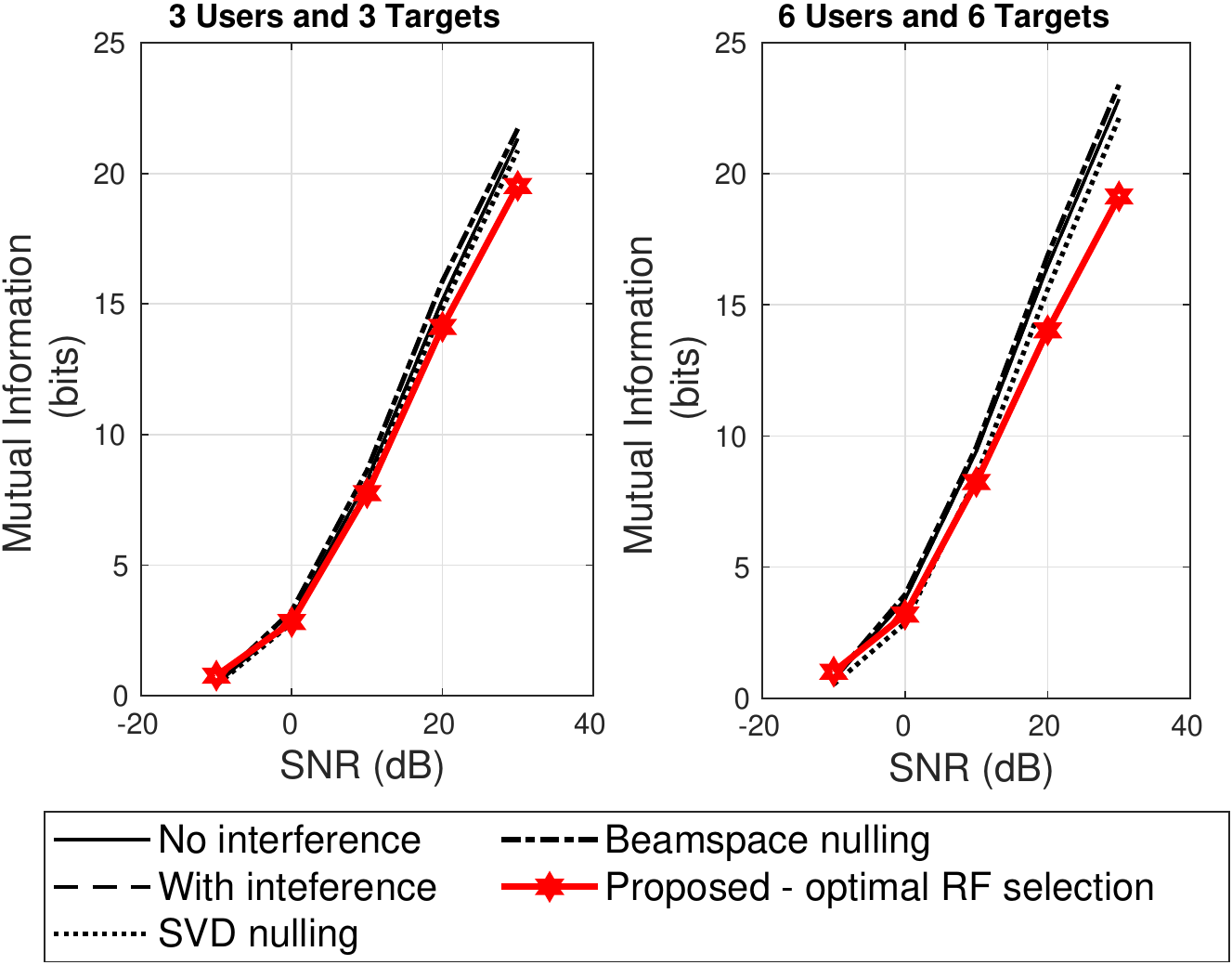}
    \caption{Mutual information performance versus SNR for different target and user scenarios, $N=32, \rho=1$.}
    \label{fig:capacityVsnr1}
\end{figure}

Let us first depict the results for the MUI over the signal-to-noise ratio (SNR), which is defined as $
    \text{SNR} = \frac{1}{\sigma_n^2}$.
In Fig. \ref{fig:capacityVsnr1}, we show the MUI for the case of $\rho=1$. Recall that, the weighting between the two MUI terms is a parameter that is being defined by the user, depending on needs of the application. Note that, for the two extreme cases, where $\rho=1$ or $\rho=0$, there is no interference, thus all techniques are expected to behave similarly. The proposed technique follows closely the other baselines, given that does not requires the instantaneous channel matrices, thus, lags some performance at the higher SNR regime.
 
\begin{figure}[t]
    \centering
    \includegraphics[scale=0.6]{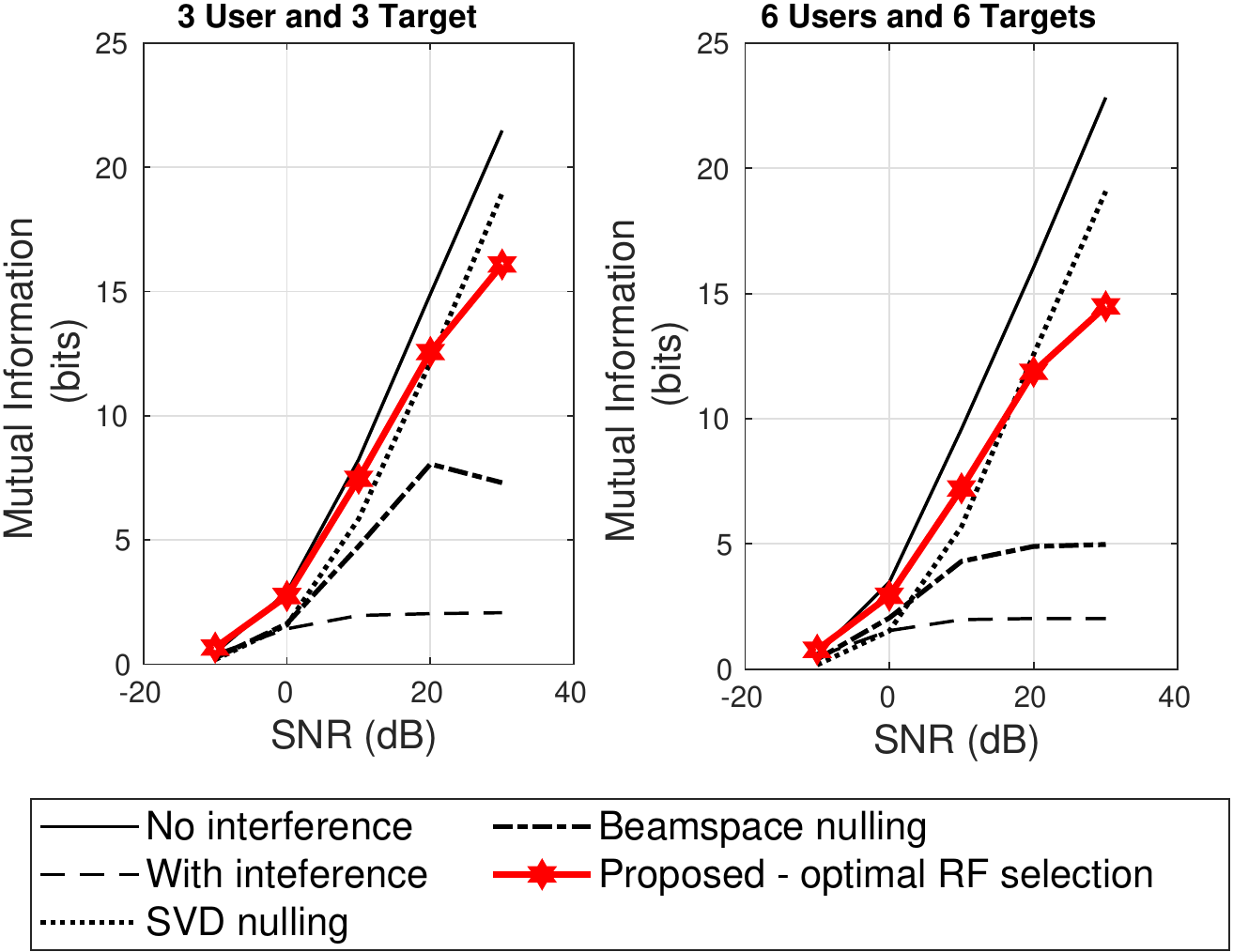}
    \caption{Mutual information performance versus SNR for different target and user scenarios, $N=32, \rho=0.5$.}
    \vspace{-5mm}
    \label{fig:capacityVsnr2}
\end{figure}

In Fig. \ref{fig:capacityVsnr2}, we show the results for the case of $\rho=0.5$, which is the case with the maximum interference. The proposed technique outperforms the other baselines for the low and medium SNR regime, thus, is proved as more robust. The performance of the SVD nulling is better for higher SNRs, while the Beamspace nulling is lagging behind the other techniques. 

\begin{figure}[t]
    \centering
    \includegraphics[scale=0.6]{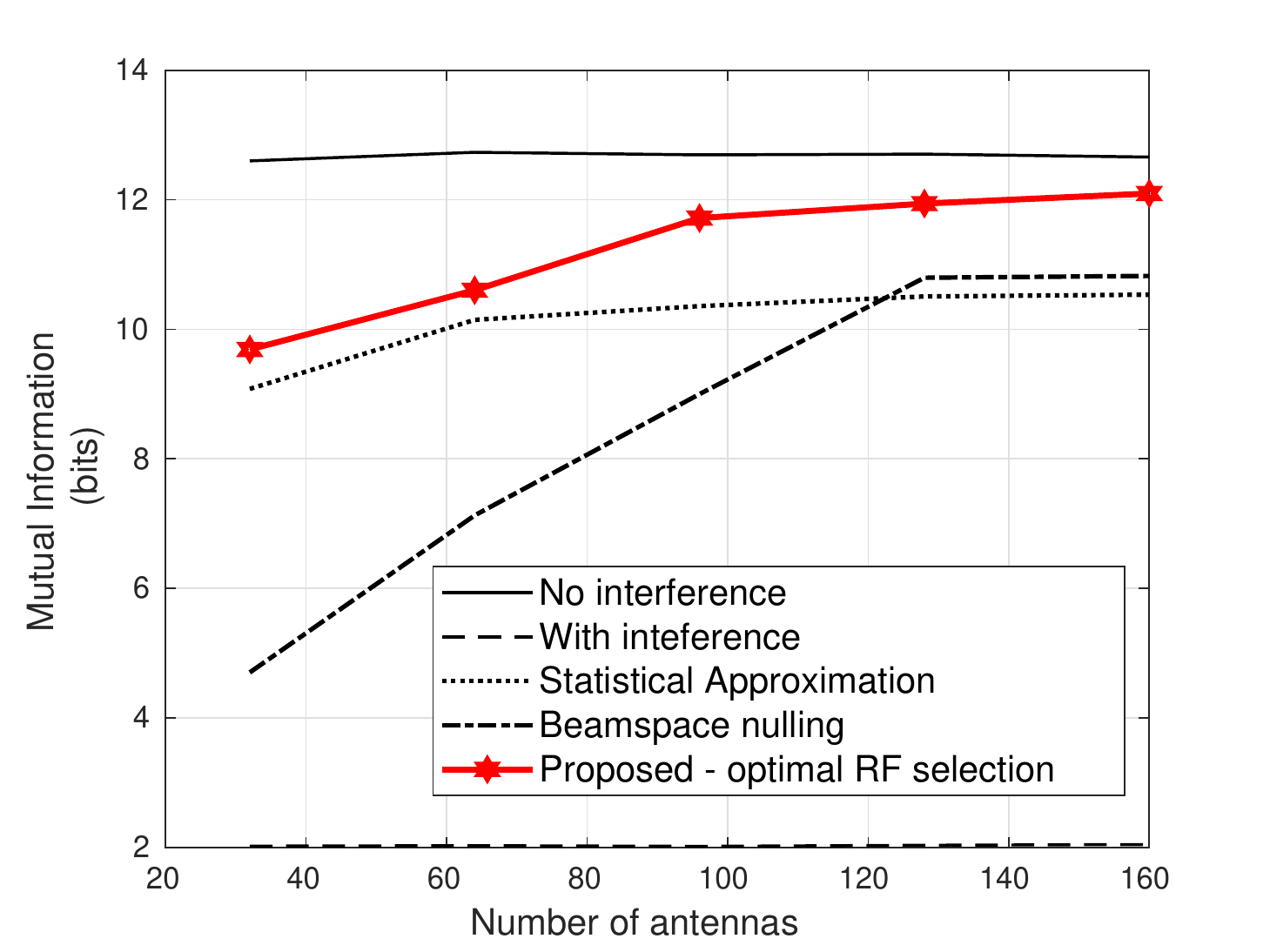}
    \caption{Mutual information performance versus number of transmit antennas at $\text{SNR}=15\text{dB}, \rho=0.5$.}
    \vspace{-5mm}
    \label{fig:capacityVsnt}
\end{figure}

In Fig. \ref{fig:capacityVsnt} we show the results for the MUI over the number of the antennas $N$, for the case of 6 users and 6 targets. As the antenna size, and thus, the angle resolution, increases, the performance of all techniques is improved. It is important to stress out that the proposed technique has significant benefits from larger antenna arrays.

\begin{figure}[t]
    \centering
    \includegraphics[scale=0.6]{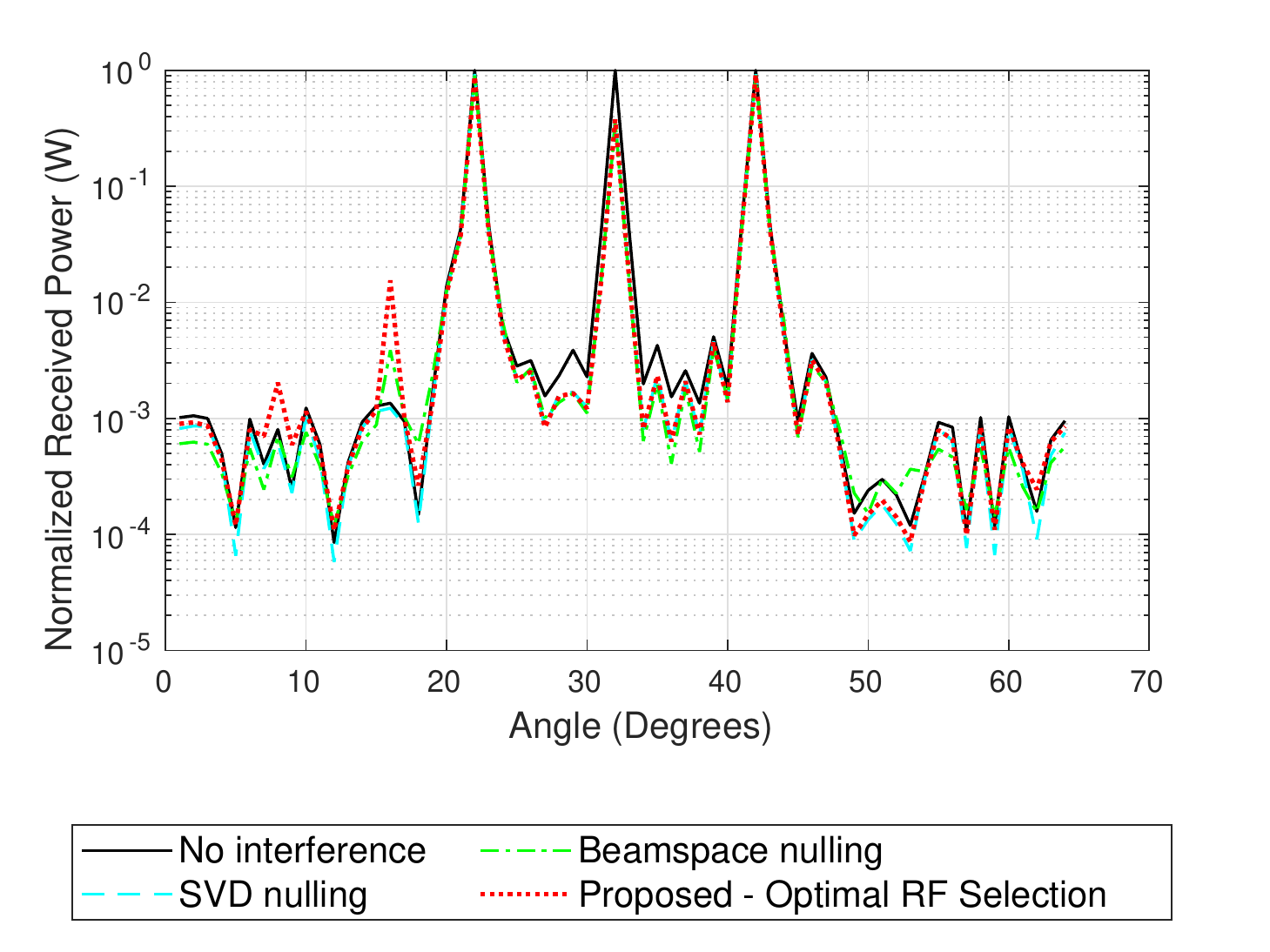}
    \caption{Normalized received power performance, $\rho=0, N=64, \text{SNR}=5\text{dB}$.}
    \label{fig:beampattern_0}
    \vspace{-1em}    
\end{figure}

\begin{figure}[t]
    \centering
    \includegraphics[scale=0.6]{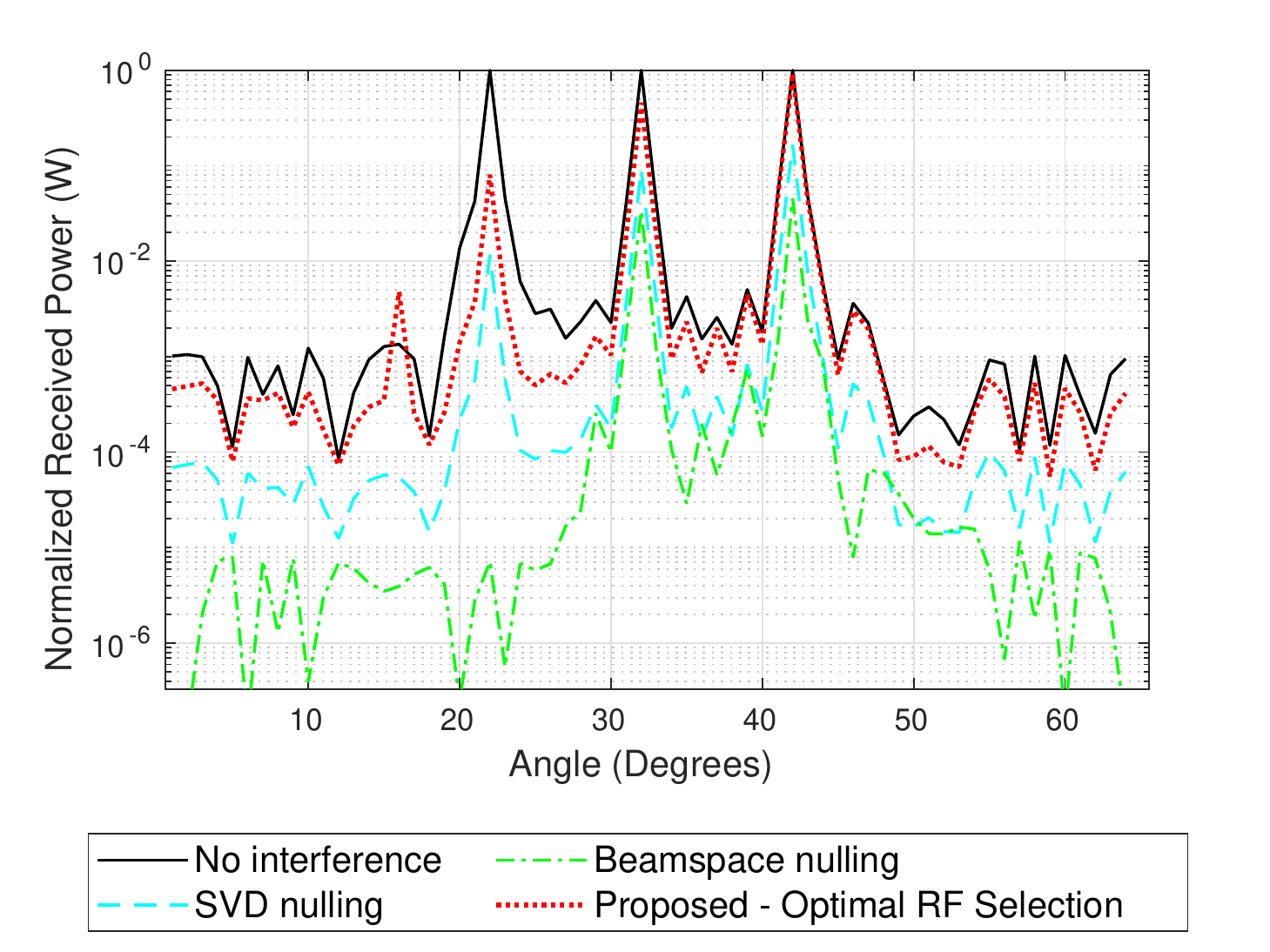}
    \caption{Normalized received power performance, $\rho=0.8, N=64, \text{SNR}=5\text{dB}$.}
    \label{fig:beampattern_08}
    \vspace{-1em}
\end{figure}

Next, we evaluate the radar performance, in terms of Normalized Received Power (NRP), defined as: 
\begin{equation}
    \text{NRP} = \frac{\Vert \mathbf{B} \mathbf{H}_\text{R} \mathbf{F}_\text{R} \Vert_F^2}{\Vert \mathbf{B}\Vert_F^2}
\end{equation}
In Figs. \ref{fig:beampattern_0} and \ref{fig:beampattern_08}, we show the beampattern for two cases of $\rho$. The radar channels have been constructed assuming that there are 3 targets placed at angles $22, 32$ and $42$ degrees.


\section{Conclusion}
This paper designs a covariance-based robust hybrid beamformer for JRC system which leads to maximization of the mutual information based on weighted dual function while taking into account the interference of one operation to the other. Fractional programming is used for obtaining optimal RF chain selection leading the flexibility in hybrid beamformer design. The proposed method shows high mutual information performance for the JRC system and favourable normalized received power when compared to the baseline cases. Different weighting factor values are also considered for performance evaluation. For instance, at SNR =15 dB and $\rho = 0.5$, we can observe that the proposed method performs better than all baseline cases except the "no interference" baseline which does not take into account the impact of interference. 

\vspace{-2mm}
\bibliographystyle{IEEEtran}

\end{document}